\documentstyle[prd,twocolumn,tighten,aps]{revtex}

\newcommand{\be}{\begin{equation}}
\newcommand{\ee}{\end{equation}}
\newcommand{\bn}{\begin{eqnarray}}
\newcommand{\en}{\end{eqnarray}}

\def\l{\label}

\newcommand{\ed}{\end{document}}

\begin{document}
\draft

\twocolumn[\hsize\textwidth\columnwidth\hsize\csname
@twocolumnfalse\endcsname

\title{On the Lorentz symmetry of the noncommutative chiral bosons}
\author{E. M.C. Abreu$^{a}$, R. Menezes$^{b}$ and C. Wotzasek$^{c}$}
\address{$\mbox{}^{a}$Instituto de F\' \i sica, Universidade
Estadual do Rio de Janeiro, 58051-970, Rio de Janeiro, Brazil\\
and Departamento de F\'{\i}sica, Instituto de Ci\^encias Exatas, Universidade Federal Rural do Rio de Janeiro\\
BR 465-07, 23851-180, Serop\'edica, Rio de Janeiro, Brazil\\
{\sf E-mail: evertonabreu@ufrrj.br}\\
$\mbox{}^{b}$Instituto de F\'\i sica, Universidade Federal da Para\'\i ba, 21945, Jo\~ao Pessoa, Brazil\\
{\sf E-mail: rms@fisica.ufpb.br}\\
$\mbox{}^{c}$Instituto de F\'\i sica, Universidade Federal do Rio de
Janeiro, 21945, Rio de Janeiro, Brazil\\
{\sf E-mail: clovis@if.ufrj.br}}

\date{\today}
\maketitle

\begin{abstract} 
We study issues of Lorentz violation symmetry in the context of the recently proposed theory of noncommutative fields \cite{CCGM}, using the soldering formalism. To this end a noncommutative chiral-boson with a deformed algebra \cite{DGMJ}, used to study these notions in D=2, is properly generalized. We verify, also for this larger group of theories that, although the structure of the Lorentz group is preserved, the velocity of light is scaled by a function of the deformation parameter, as recently claimed.  However, we found a sub-set of models where the velocity of propagation is maintained in spite of the presence of the deformed algebra. Effects of a preferred-frame of reference manifest by the presence of birefringence were also studied in the chiral boson framework leading to the scalar sector of the extended Standard Model recently proposed.
\end{abstract}

\vskip2pc]



\section{Introduction}

Last year a quantum theory of noncommutative fields was elaborated as a generalization of noncommutative quantum mechanics \cite{CCGM} which is rather different from the usual quantum field theory over a canonical noncommutative
spacetime \cite{string}. These results seem relevant for certain classes of observations in astrophysics \cite{astrobservation}, such as the lack of symmetry between particles and antiparticles, which could play a role in describing the observed matter asymmetry in the Universe.
The theory although formulated in classical (commutative) spacetime admits
nonzero equal-time commutation relations between the basic fields.
What seems important for us in this context is the possibility to make a phenomenological use of such formulation to study deviations from Lorentz symmetry \cite{DGMJ}. 

Lorentz invariance is one of the cornerstones of modern quantum field theory.
This is a symmetry respected by the Standard Model of known elementary particles and their interactions \cite{CK} but the exact character of physics beyond it remains an open question.
Possible signals of Lorentz violation could, therefore, be indicative of new physics, e.g. quantum gravity at the Planck scale \cite{quantumgravity}. 

It is however quite hard to formulate a theory to describe physical phenomena at such small scale.
The impossibility to satisfactorily formulate gravity as a relativistic quantum field theory is behind this difficulty. Since some of the basic assumptions of modern quantum field theory might fail at high-energy there seems to be no a priori reason why such schemes should be the correct framework to formulate these questions. As recent developments
in quantum gravity suggest, at very high energies Lorentz invariance could not even be an exact symmetry.

While all present theoretical suggestions of Lorentz violation could not be considered as predictive, there is nevertheless great interest in the possibility that Lorentz violation induced by Planck scale physics could offer an observational window into quantum gravity.
It has been argued recently  \cite{quantumgravity} that remnants of quantum gravity could be seen from dispersion relations violating Lorentz invariance.
It is clear however that any theory of particle physics at very-high energy should reduce to a known quantum field theory at low energies.
Lorentz invariance is then seen as a good low-energy symmetry which may be violated at very high energies \cite{highenergy}.
Since our low-energy theories are relativistic quantum field theories, it is interesting to explore possible
extensions of the QFT framework which could produce departures from exact Lorentz invariance.
For example, in the presence of certain forms of Lorentz violation, light propagating
through the vacuum will experience birefringence.  

It might be worth adding departures from Lorentz symmetry by non-quantum-gravity mechanism.
In the random dynamics programme \cite{Chadha:1982qq}, it was proposed that the various symmetries observed in nature could be regarded as infrared attractive fixed points of a large class of theories which are not endowed with these symmetries a priori while in \cite{np}, the simplest Lorentz non-invariant coupling in QFT in order to study possible deviations from the Lorentz invariant laws of nature was invoked. In \cite{klinkhamer}, the cylindrical topology of the universe was explored.  The so-called CPT anomaly, that depends on the global structure (topology) of space, and occurs for a class of chiral gauge theories that includes the standard model of elementary particle physics was then exploited.  More recently, Kostelecky, Lehnert and Perry analyzed the association of the spacetime-varying couplings with Lorentz and CPT violation with a general effect in theories with derivative couplings to cosmological fields \cite{klp}.

Motivated by the appearance of noncommutative spaces in string theory, there has been recently
much activity in an area called as noncommutative quantum field theory \cite{string}.
The non-commutative theories present examples
that violate relativistic invariance \cite{qednc,VSL} and modify, in a peculiar way, the short-distance behavior
of the theory.
In \cite{CCGM} it has been proposed an extension of the relativistic quantum field theory framework based on the notion of what was called as a {\it noncommutative field} (NC). 
In particular, the assumption that noncommutativity in the field space of a QFT produces Lorentz-violating dispersion
relations has been investigated.
They found, for instance, that cosmic ray physics is sensitive to a NC scale as low as the Planck length.

In a recent paper the noncommutative field space formulation was used to analyze the Abelian bosonization for
a two dimensional system \cite{DGMJ}.  
The rationale here is that an analysis in a $D=2$ spacetime theory can be useful in disclosing the basic physics underlying this problem.
They found that for chiral bosons in a non-commutative field space conformal invariance continues to hold and that
the non-commutativity in the field space leads to free fermions when chiral bosons are fermionized.

However Lorentz symmetry violation due to noncommutativity is not yet established in the context of \cite{CCGM} as compared to the usual formulation of field theory over canonical noncommutative spacetime simply because we still lack explicit model realizations. In this Report, the connection between Lorentz invariance violation and noncommutativity of fields
in a quantum field theory of chiral bosons is resumed.
In the next Section a generalized model of non-commutative field space chiral bosons with a real one-parameter deformed symplectic algebra \cite{DGMJ} is investigated upon the soldering of the individual chiralities.
It is shown that, for a large class of chiral theories generalizing \cite{DGMJ}, the structure of the Lorentz group is preserved, despite the assumption that noncommutativity in the field space of a QFT produces Lorentz-violating dispersion relations. The original Lorentz asymmetry then becomes manifest as a scaling of the velocity of light. Such a result, corroborating the findings of \cite{DGMJ}, becomes manifest by the soldering formalism leading to a conventional scalar field structure for a whole family of NC-chiral bosons with a deformed symplectic structure. However, what is even more surprising is that there is a sub-class of theories where, despite the NC deformation, the velocity of light is preserved. We shall disclose the physics behind such structures.

Birefringence is commonly found in conventional electrodynamics in the presence of anisotropic media.
The birefringence of the vacuum stands among the unconventional properties of radiation arising from Lorentz violation. 
In Section III we shall consider issues of birefringence leading to possible extensions of the previous findings. 
Here we shall consider these notions upon the soldering of noncommutative chiral fields when the vacuum displays the birefringence phenomenon. 
It is manifest here in the form of chiral components with different propagation velocity which, upon soldering, leads to explicitly Lorentz violating forms.

\bigskip

\section{Soldering of NC chiral bosons}

We begin with a related form of the action for the noncommutative chiral boson proposed in \cite{DGMJ} (DGML) where the noncommutativity is manifested in the fields \cite{CCGM},
\bn
\l{1}
{\cal L}_0 &=& a(\theta)\,(\dot{\varphi}_+\,{\varphi'}_+ \,-\,\dot{\varphi}_-\,{\varphi'}_-) \nonumber\\
&+& 2\,d(\theta)\,\dot{\varphi}_+\,{\varphi'}_-\,-\,{\varphi'}^2_+\,-\,{\varphi'}^2_- \; ,
\en
where use of natural units is maintained as usual. Here $a(\theta)$ and $d(\theta)$ are dimensionless functions of the noncommutative parameter $\theta$. 
Let us recall that in the NC-field approach $\theta$ is a dimensionless parameter as well. The noncommutative deformation is parametrized by $d(\theta)$ while $a(\theta)$ normalizes the symplectic structure of the original chiral boson theories. Together they fix the velocity of propagation of the chiral bosons. They can, in principle, be determined, as in the original DGML action but we will keep them arbitrary for awhile aside from the fact that when $\theta\to 0$ they should have a smooth limit as $a(\theta)\to a(0)=\mbox{const.}$ and $d(\theta)\to 0$, being usual to consider the limiting value as $a(0)=1$.  In such limit the chiral components decouple and travel with the ordinary light velocity 
\be
c(\theta) \to c_0^{(\pm)} = \frac {\pm 1}{a(0)}\, .
\ee
We will see that they play a very important role in the construction of different sets of physical solutions that will allow us to perform a precise analysis of the Lorentz invariance.  Meanwhile, let us consider these functions as ``undetermined coefficients" to be fixed by the application of the soldering formalism.
As notational remark, observe that dots and primes mean the usual time and space derivatives.

The soldering formalism \cite{solda}, is an iterative method that permits us to construct an effective action invariant under a specific gauge symmetry disclosing interesting physical features analogous to the interference phenomena.  The method uses primarily the Noether gauging procedure that helps us to fuse together the variables representing the different aspects of a theory.   It works by elevating a global symmetry to a local form thanks to a mutual cancellation of the obstruction to gauge symmetry of the individual components. It is the gauge field introduced in the process the agent responsible for this new symmetry and may be eliminated after the gauging is complete. In this Letter we will explore just the main steps of the method.  For a detailed reading, see \cite{capitulo}.

Chiral bosons have been presented in different physical context \cite{siegel,florjack}. It has been shown \cite{abreuwotz} that the first-order, noninvariant Floreanini-Jackiw formulation displays the same dynamical content of the Siegel formulation while its chiral diffeomorphism is realized by another field called as noton \cite{hull}. The formulation proposed in \cite{florjack}, which is the one adopted in \cite{DGMJ} to examine issues of Lorentz symmetry violation, is not Lorentz covariant. However the equations of motion were shown to be invariant, a property that is expected to be broken by the NC-deformation introduced in \cite{DGMJ}.  While it is possible to gauge the global vectorial symmetry of the model proposed in \cite{florjack}, the gauging of the global chiral symmetry will alas fail. However the good news is that the obstruction to the gauging is not field dependent.  It is therefore possible to adjust the right and the left modes to obtain a local symmetry, not for the components but for the composite. It is this local symmetry that allows the fusion of the individual chiral components into a new ``multiplet".

The soldering formalism is the well suited algorithm for fusing together opposite aspects of a global symmetry, such as the right and left propagating modes of chiral bosons just discussed.  However, due to the interaction between chiralities introduced by the noncommutative deformation in the DGML model, a deformation in the soldering should as well be expected. Besides, the need for a local symmetry should impose strong restrictions over the soldering parameters with dramatical consequences. Let us then gauge the following global symmetry obeyed by the chiral modes in ({\ref{1}),
\be
\l{2}
\delta \varphi_{\pm}\,=\, \lambda_{\pm} \,\alpha\;\;, 
\ee
where $\alpha$ is the gauge parameter and the functions $\lambda_{\pm}= \lambda_{\pm}(\theta)$ shall give a parametrization of the NC-deformation. The gauging of the global parameter $\alpha \to \alpha(x,t)$ will be done in the sense of the soldering. As so, we will refer to the local version of (\ref{2}) as the soldering symmetry.  We will see that the $\lambda_{\pm}$ parameters will help us to construct the effective field of the final action. 
With this symmetry, a simple calculation shows that,
\be
\label{deltalzero}
\delta\,{\cal L}_0 \,=\,\alpha'\,{\cal J}\, ,
\ee
where ${\cal J}$ is the Noether chiral current given by
\bn
\l{3}
{\cal J}&=& 2\,[\,(a\,\lambda_+\,+\,d\,\lambda_-)\,\dot{\varphi}_+\,-\,(a\,\lambda_-\,-\,d\,\lambda_+)\,\dot{\varphi}_- \nonumber\\
&-& (\lambda_+\,{\varphi'}_+\,+\,\lambda_-\,{\varphi'}_-)\,]\;\;,
\en
that parametrizes the lack of gauge symmetry of the original action ${\cal L}_0 $, as shown in (\ref{deltalzero}).

The next step is to introduce an auxiliary field that helps in the gauging procedure.  Let us represent this field by $B$, known as the soldering field.  Now we can construct the first-iterated correction of the DGML action (\ref{1}) as
\be 
\l{4}
{\cal L}_1\,=\,{\cal L}_0\,-\, B\,{\cal J}\;\;.
\ee
Notice that we are looking for an action which is gauge invariant under (\ref{2}).

This objective was not accomplished yet since,
\bn
\l{5}
\delta {\cal L}_1&=&-\,B\,\delta{\cal J}  \nonumber \\
&=& -2 B[\lambda_+(a\lambda_+ + d\lambda_-)\dot{\alpha} \\
&-&\lambda_-(a\lambda_- - d \lambda_+)\dot{\alpha} 
- ({\lambda}^2_+ + {\lambda}^2_-){\alpha'}]\nonumber
\en
where we have chosen $\delta B\,=\,\alpha'$ to cancel (\ref{deltalzero}) and used (\ref{2}) once again. For the conventional theory of chiral bosons where the deformation function $d(\theta)$ vanishes (and $\lambda_\pm\to 1$) a gauge invariant action would follow at this stage since the last term in the r.h.s. of (\ref{5}) is integrable. However, the analogous expression here,
\be
\l{8}
{\cal L}_{2}\,=\,{\cal L}_0\,-\,B\,{\cal J}\,-\,({\lambda}^2_+\,\,+\,{\lambda}^2_-)\,B^2\;\;,
\ee
fails to be gauge invariant because of the presence of the NC deformation
\bn
\l{D2}
\delta {\cal L}_2&=& -2 \,[\lambda_+(a\lambda_+ + d\lambda_-) \nonumber\\
&-&\lambda_-(a\lambda_- - d \lambda_+)]\, B\,\dot{\alpha}\, .
\en
It is at this point that the presence of the soldering parameters play an important role. With a suitable choice of these parameters we may still achieve a gauge invariant action allowing for the completion of the soldering. From (\ref{D2}) we can see that this goal may be accomplished as long as the following constraint is satisfied
\be
\l{6}
a\,({\lambda}^2_+\,\,-\,{\lambda}^2_-)\,+\,2\,\lambda_+\,\lambda_-\,d\,=\,0\;\;.
\ee
With this restriction in (\ref{D2}) we can write down that,
\be
\l{7}
\delta{\cal L}_2\,=\,0
\ee
to finally get the desired gauge invariant action.

Eliminating the $B$ field through its equations of motion and substituting it back in (\ref{8}) we obtain an effective action
\be
\l{9a}
{\cal L}_{eff}\,=\,{\cal L}_0\,+\, {1 \over 8}\,{\cal J}^2\;\;,
\ee
that, after using the Noether current (\ref{3}), is able to produce the desired soldered action.

For this, we have to go back to relation (\ref{6}) that has the following solution,
\be
\l{9}
\lambda_{\pm}\,=\,\sqrt{1\,\mp\,d\,c}\, .
\ee
Notice that when we take the smooth limit to the commutative case, $d(\theta)\to 0$ then $\lambda_\pm\to 1$ as expected.
Here $c$ is a new parameter that can be written as
\be
\l{10}
{1 \over c}\,=\,\sqrt{a^2\,+\,d^2}\;\; ,
\ee
that will play distinctive role in the sequel.
It is important to observe at this juncture that both the soldering parameters $\lambda_\pm$ and $c$ are {\it theory dependent}, taking different values for different deformations.

With this solution in mind we can construct the following relations involving the $\lambda_\pm$ parameters,
\bn
{\lambda}^2_+\,\,+\,{\lambda}^2_-&=& 2\, , \\
{\lambda}^2_+\,\,-\,{\lambda}^2_- &=& -\,2\,d\,c\, , \\
\lambda_+\,\lambda_- &=& a\,c\, , \\
a\,\lambda_-\,-\,d\,\lambda_+ &=& {\lambda_+ \over c}\, , \\
a\,\lambda_-\,+\,d\,\lambda_+ &=& {\lambda_- \over c} \, ,
\en
where the first relation is an imposed normalization condition.
Using this solution we can rewrite the Noether current (\ref{3}) as
\bn
{\cal J}\,&=&\, 2\,[{1 \over c}\,(\lambda_-\,\dot{\varphi}_+\, - \,\lambda_+\,\dot{\varphi}_-)  \nonumber\\
&\,-&\,(\lambda_+\,{\varphi'}_+\,+\,\lambda_-\,{\varphi'}_-)]\;\;.
\en
Next, let us define a new composite scalar field
\be
\Phi\,=\,\lambda_+\,\varphi_-\,-\,\lambda_-\,\varphi_+\;\; ,
\ee
in which the chiral boson components are pondered by the soldering parameters $\lambda_\pm$ that compensates for the interaction introduced into the symplectic sector.
Substituting all these results in (\ref{9a}) it is then easy to see that,
\bn
\label{scalar}
{\cal L}_{eff}\,&=&\,{1 \over 2c^2}\,{\dot{\Phi}}^2\,-\,{1\over2}\,{\Phi'}^2 \nonumber\\
&=& \frac 12 \,\partial_\mu \Phi \,\partial^\mu \Phi\;\;,
\en
where $x_\mu = (ct,x)$.
It is important to note that this result, appearing from the fusion (soldering) of noncommutative chiral bosons, leads to a Lorentz invariant structure.  However we can see that the light velocity has now a $\theta$--dependent value, given by (\ref{10}).
The new velocity depends clearly on the choice of deformation adopted. Adopting the parameters in the action (\ref{1}) from the actual DGML action \cite{DGMJ}
\bn
a&=& \frac{1}{1\,+\,\theta^2}\, , \\
d&=&-\,\frac{\theta}{1\,+\,\theta^2}\, ,
\en
we obtain that
\be
c\,=\,\sqrt{1\,+\theta^2}\, \, ,
\ee
where the noncommutative parameter is usually assumed $\theta \ll 1$. 
For this case the soldered scalar function assumes a quite interesting form given by
\be
\Phi = \frac 1{c}\left[\sqrt{1 + \frac \theta{c}}\, \varphi_- - \sqrt{1 - \frac \theta{c}}\,\varphi_+\right]\, .
\ee
This value departs from the usual value for the velocity of light (in natural units $c=1$) and corroborates the findings of DGML for the new value for the velocity of light.

At this juncture, it is interesting to clarify if the velocity of propagation for the composite is the result of a deformation introduced by the symplectic interaction and deviates from the velocity of the chiral components. To compare the propagation velocity of the composite scalar field $\Phi$ with the correspond propagation velocity of the components we rewrite (\ref{1}) in matrix form as,
\be
{\cal L} = \dot\varphi_k M_{km} \varphi_m ' - {\varphi_k '}^2\, ,
\ee
where
\bn
\l{matriz}
\left(M\right) &=& \pmatrix{a(\theta)  & d(\theta) \cr
	d({\theta})  & {-a(\theta)} \cr} \, ,
	\en
and compute their equations of motion. Here $k,m=\pm$ and the matrix satisfy
\bn
\left(M^2\right)_{km} &=& \left(a^2 + d^2\right) \delta_{km}\, ,\nonumber\\
M^{-1} &=& \frac 1{a^2 + d^2}  M\, .
\en
{}From here the equations of motion follow straightforwardly,
\be
\dot\varphi_k = \left( M^{-1}\right)_{km} \varphi_m '\, ,
\ee
where one constant of integration has already been fixed, and
\be
\ddot\varphi_k = \frac 1{a^2 + d^2}\, \varphi_k '' \, ,
\ee
which shows that the propagation velocity for the chiral components are equal and identical to the propagation velocity of the composite.
Therefore, since all three waves propagate at the very same velocity, the fact that the scalar composite's propagation velocity deviates from the ordinary light velocity just reflects the fact that the $\theta$--deformation modifies the velocity of the chiral components.

An important feature of the soldered solution just obtained is that it fits the generalized action (\ref{1}) where the parameters are still free. Therefore, if one insists in preserving the usual $c=1$ value even in the presence of the deformation $ \forall \,\theta$, one finds from (\ref{10}) or
\be
a^2\,+\,d^2\,=\,1\;\;,
\ee
that there is a whole one-parameter family of theories whose propagation velocity remains invariant, equal to the ordinary light velocity despite the presence of the $\theta$--deformation. For example, let us consider the specific theory given by
\be
\l{unitary}
a\,=\,\frac{1}{\sqrt{1+\theta^2}} \qquad \mbox{and}\qquad d\,=\,\frac{\theta}{\sqrt{1+\theta^2}}\, ,
\ee
leading to unitary velocity of light.
This is an unexpected and extraordinary result put forward by the soldering of the NC-deformed chiral theories.
In fact, keeping the basic structure proposed in DGML for the generalized theory,
\be
\l{estrutura}
a(\theta)=\frac 1{f(\theta)} \,\,\, ; \,\,\, d(\theta) = \frac\theta{f(\theta)}\, ,
\ee
with the deformation encoded in the $f(\theta)$ function, leads us to a new velocity of light as,
\be
c\,=\,\frac{f(\theta)}{\sqrt{1\,+\theta^2}}\, .
\ee
It should be noticed that for such a class of theories there is a unique answer leading to the usual $c=1$ case that is \be
\l{funcao}
f(\theta)= \sqrt{1\,+\theta^2}\, ,
\ee
which is the result already supplied in (\ref{unitary}).

In the final part of this Section we shall discuss a second approach the DGML theory trying to disclose the special physics behind the NC-deformation proposed there. To this end let us recall that a single chiral boson $\varphi_\pm$ may be obtained from the theory of a scalar field $\phi$ by applying the chiral constraint $\pi = \pm \phi '$ in phase-space. To study the case at hand one may try to start with a couple of scalar fields and find a generalized chiral constraint that includes the $\theta$--deformation. Let us then write the action of two scalar fields, in phase space as,
\be
{\cal L}=\pi_k \dot\phi_k - \frac 12 \pi_k^2 - \frac {v^2}2 {\phi_k '}^2\,\,\, ; k=1,2
\ee
where the dimensionless parameter $v$ represents the propagation velocity as usual.
Next we try to obtain the deformed chiral boson structure with a suitable restriction over the phase-space variables as,
\be
\l{constraint}
\pi_k = M_{km} \phi_m '\, .
\ee
Clearly the choices $M_{km} = \pm \delta_{km}$ and  $M_{km} = \pm (\sigma_3)_{km}$ will restrict the original theory to a couple of chiral bosons with the same chirality in the first case and with opposite chiralities in the second case.
To obtain the deformed symplectic structure of (\ref{1}) we adopt the matrix already defined in (\ref{matriz}).  Substitution of the generalized constraint (\ref{constraint}) into the Hamiltonian density gives
\bn
{\cal H} &=& \frac 12 \phi_k ' \left(v^2\, \delta_{km} + M_{kl}M_{lm}\right)\phi_m '\nonumber\\
&=& \frac 12 \left(v^2 + a^2 + d^2\right) {\phi_k '}^2\, .
\en
It is now clear that in this approach an extra constraint must be imposed with drastic consequences.  In order to reobtain the Hamiltonian structure of (\ref{1}) the parameters must satisfy the condition
\be
\l{restricao}
\left(v^2 + a^2 + d^2\right) = 2\, .
\ee
This establishes a relation between the propagation velocity of the scalars fields and the propagation velocity of the chiral components, (Eq.\ref{10}).  If we allow for the  possibility that they could be distinct, then 
\be
v(\theta) = \sqrt{2 - a^2 - d^2}\, ,
\ee
which generalizes the findings in DGML.
However, if the (physically) more plausible choice that $v = c$ is adopted then (\ref{restricao}) can be rewritten as
\be
x^2 - 2x +1 = 0\;\; , \;\;  x=c^2=v^2
\ee
leading to a unique solution for this case as $c=1$. In such situations, therefore, the use of (generalized) chiral constraints are not able to reproduce the whole class of deformed chiral bosons as proposed in DGML but only the restricted sub-class with unitary light velocity. This result corroborates the previous findings regarding the class of theories that, despite the NC-deformation, preserve the light velocity. 

As a final remark, we observe that adding and subtracting a mix space-time derivative term ($\dot\Phi\,\Phi '$) into (\ref{scalar}) brings the composite scalar field into its conventional form at the price of including an extra, explicitly Lorentz violating term but still the vacuum is not birefringent. This situation will be clarified in the next section when the symmetry of the vacuum will be explicitly broken.

\section{Birefringence}

Recently general Lorentz-violating extensions have been constructed for the Standard Model. 
They consist of the minimal Standard Model plus small Lorentz (and also CPT) violating
terms. Such extensions have provided a theoretical framework
for many searches for Lorentz and CPT violations.
In practice, one often works with a particular limiting theory extracted from the
Standard-Model extension. For example, a Lorentz-violating modified electrodynamics is usually extracted from the photon sector of the extended Standard Model.
The modified electrodynamics maintains the usual gauge invariance, is covariant
under observer Lorentz transformations and includes both CPT-even and -odd terms.
The theory predicts several peculiar features that lead to sensitive tests of Lorentz symmetry.
More important for us here is that in the presence of certain forms of Lorentz violation, light propagating
through the vacuum will experience birefringence. 

In this Section we shall consider the consequences of birefringence in the vacuum, in an extension of the D=2 model of noncommutative fields studied in the preceding section. It will be clear next that some extensions of the theory (\ref{1}) are able to produce such effect. 

\subsection{Non-deformed chiral bosons}

For simplicity, let us consider first a model for chiral bosons without the deformation piece as,
\bn
\l{2.1}
{\cal L}_0 &=& a_+(\theta)\,\dot{\varphi}_+\,{\varphi'}_+ \,-\, a_-(\theta)\,\dot{\varphi}_-\,{\varphi'}_- \nonumber\\
&-&\,{\varphi'}^2_+\,-\,{\varphi'}^2_- \; .
\en
The velocity of propagation for the chiral components are determined by the coefficients of the symplectic terms as,
\be
\label{cvelocity}
c_\pm = \pm\frac 1{a_\pm(\theta)}\, ,
\ee
which are parametrized by $\theta \in \Re$. Since $c_+ \neq c_-$ the system displays birefringence.

To implement the soldering we follow the earlier procedure and gauge the global symmetry (\ref{2}). The lack of gauge symmetry presented by the zeroth-order action, Eq.(\ref{deltalzero})
may be compensated with the introduction of a gauging field $B$, transforming as $\delta B= \alpha\, '$, and the Noether current,
\be
\label{current}
J = 2\left[\lambda_+\left(a_+ \dot\varphi_+ -\varphi_+ '\right)
- \lambda_-\left(a_- \dot\varphi_- +\varphi_- '\right)\right]\, .
\ee
Following the gauging procedure, we find that the second-iterated action,
\be
{\cal L}_{2}\,=\,{\cal L}_0\,-\,B\,J\,-\,({\lambda}^2_+\,\,+\,{\lambda}^2_-)\,B^2\;\;,
\ee
becomes gauge-invariant if the following condition over the soldering parameters holds,
\be
\l{condition2}
a_+\, \lambda_+^2 - a_-\, \lambda_-^2 =0\, .
\ee
This constraint may be written as a pair of conditions which, in matrix form read
\be
\pmatrix{a_+(\theta)  & {-1} \cr
	{1}  & {-a_-(\theta)} \cr}\pmatrix{\lambda_+ \cr -\lambda_- \cr} = 0\, .
\ee
A nontrivial solution for $\vec \lambda$ demands 
\be
a_+(\theta) \, a_-(\theta) = 1\,.
\ee
For simplicity, let us introduce a normalization condition as
\be
\l{normalization}
\lambda_+^2 + \lambda_-^2 = 2 \, .
\ee
The one-parameter solution for these conditions is given in terms of $\theta$ as,
\be
a_\pm(\theta) = \frac 1{r \mp \theta}\, ,
\ee
and
\be
\lambda_\pm(\theta) = \sqrt{\frac{r\mp \theta}{r}}\, ,
\ee
where
\be
r=\sqrt{1+\theta^2}\, .
\ee
Elimination of the soldering field $B$ leads to an effective action as
\be
{\cal L}_{eff}\,=\,{\cal L}_0\,+\, {1 \over 8}\,{ J}^2\, .
\ee
This effective action may be cast in a more interesting form with the introduction of the soldered doublet
\be
\Phi = \lambda_- \varphi_+ - \lambda_+ \varphi_-\, .
\ee
Indeed, using (\ref{current}) together with solution for the soldering parameters above, the soldered action becomes
\be
\label{kostelecky}
{\cal L}_{eff} = \frac 12 \partial_\mu \Phi \,\partial^\mu\Phi + \frac 12 K_{\mu\nu}\partial^\mu \Phi \,\partial^\nu\Phi\, ,
\ee
where the symmetric and traceless Kostelecky matrix is
\be \l{3.1a}
\left(K_{\mu\nu}\right) = \pmatrix{0 & \theta \cr \theta & 0 \cr}\,.
\ee
Therefore, we see that considering the (Lorentz symmetry violating) situation where the vacuum of the $D=2$ chiral bosons presents birefringence, leads to a model for scalar fields (without noncommutativity) with an explicit Lorentz violating term \cite{scalar}.  The effects of the birefringence are all encoded in the Kostelecky term parametrized by $\theta$ in (\ref{kostelecky}).


\subsection{Coordinate transformations and Lorentz invariance}

Before considering the sympletic deformation, let us digress on the meaning of the action (\ref{kostelecky}). 
In this respect it seems worth recalling that occasionally Lorentz breaking is connected to the loss of coordinate independence.  Therefore, we have to analyze some arguments behind this underlying principle before describing its consequences for dispersion relations \cite{lehnert}.

For non-interacting fields and certain other systems, Lorentz symmetry breaking can sometimes be removed from the theory by a convenient sequence of transformations and field redefinitions.
To see this let us observe that the action (\ref{kostelecky}) can be rewritten as,
\bn
\label{kostelecky2}
{\cal L}_{eff} &=& \frac 12\, g_{\mu\nu}\,\partial^\mu \Phi \,\partial^\nu\Phi + \frac 12 K_{\mu\nu}\partial^\mu \Phi \,\partial^\nu\Phi\, \nonumber \\
&=& \frac 12\,\left(\,g\,+\, K\,\right)_{\mu\nu}\,\partial^\mu \Phi \,\partial^\nu\Phi
\en
where the tensor $(g\,+\,K)_{\mu\nu}$
\be \l{3.2}
\left(\,g\,+\,K\right)_{\mu\nu} = \pmatrix{1 & \theta \cr \theta & -1 \cr}\,\,.
\ee
acts as an effective metric whose off-diagonal pieces, i.e., the $\theta$-factor, can be interpreted as an improper choice of coordinates.  In fact, $(g\,+\,K)_{\mu\nu}$ can be diagonalized by solving its charecteristic equation
\be
\left| \begin{array}{cc}
\alpha-1 & -\theta \\ 
-\theta & \alpha+1  \end{array} \right| \,=\,0\;\;.
\ee
From the eigenvalues,
\be
\alpha_{\pm}\,=\,\pm\,\sqrt{1+\theta^2}\;\;,
\ee
we construct an orthogonal matrix $R$, 
\be
R\,=\,\frac 1{\sqrt{2}}\,\pmatrix{\frac{\delta_+}{\sqrt{\theta^2 + \delta_+}} & \frac{\delta_-}{\sqrt{\theta^2 + \delta_-}} \cr \frac{\theta}{\sqrt{\theta^2 + \delta_+}} & \frac{\theta}{\sqrt{\theta^2 + \delta_-}} \cr}\,.
\ee
where $\delta_\pm\,=\,1 \pm \sqrt{1+\theta^2}$.


Making a transformation of coordinates such that
\bn
x_{\mu}'&=&R\,{x}_\mu \nonumber 
\en
we obtain
\bn
t'&=&{\frac{\delta_+}{\sqrt{\theta^2 + \delta_+}}}\,t\,+\,{\frac{\delta_-}{\sqrt{\theta^2 + \delta_-}}}\,x \\
x'&=&\frac{\theta}{\sqrt{\theta^2 + \delta_+}}\,t\,+\,\frac{\theta}{\sqrt{\theta^2 + \delta_-}}\,x\;\;.
\en
Next, the coordinates $x_{\mu}'$  can be rescaled to absorb the eigenvalue factor, so that the Lagrangian (\ref{kostelecky}) is canonically equivalent to a Lorentz invariant theory. Therefore,
we see, after a simple calculation, that the metric in these new coordinates is Minkowskian.


It can be shown that this argument also applies to models with multiple fields that exhibit the same Lorentz violation. However, when the model has additional fields with different Lorentz violation, the argument is no longer valid.  If we perform the transformation above, it only makes the Lorentz breaking go to another sector of the theory.  Although it is possible to pick up coordinates so that each particle of the model propagates conventionally, the Lorentz violation cannot be eliminated simultaneously from every sector \cite{km}.


\subsection{Deformed chiral bosons}

Next we shall include the effects of the noncommutative deformation proposed in \cite{DGMJ}. Let us then consider the following theory
\bn
\l{3.1}
{\cal L}_0 &=& a_+(\theta)\,\dot{\varphi}_+\,{\varphi'}_+ \,-\, a_-(\theta)\,\dot{\varphi}_-\,{\varphi'}_- \nonumber\\
&+& 2 d(\theta)\, \dot\varphi_+\,\varphi_-' \, -\,{\varphi'}^2_+\,-\,{\varphi'}^2_- \; .
\en
To determine the velocity of propagation for the chiral components we analyse the dispersion relation of this theory. The first-order field equations, reading
\bn
\l{fieldequations2}
a_+\, \dot\varphi_+' -\, \varphi_+'' +\, d \, \dot\varphi_-'&=& 0 \, ,\nonumber\\
-a_-\, \dot\varphi_-' -\, \varphi_-'' +\, d \, \dot\varphi_+'&=& 0\, ,
\en
lead to a pair of decoupled second-order equations,
\be
\beta^2\, \ddot\varphi_\pm +\,\left(a_+ - a_-\right) \dot\varphi_\pm' -\, \varphi_\pm''= 0\, ,
\ee
where
\be
\l{beta}
\beta = \sqrt{d^2 + a_+ a_-}\, .
\ee
These equations admit plane-wave solutions,
\be
\varphi_\pm = \exp\left(i\omega_\pm t - i k_\pm x\right)\, ,
\ee
leading to the following dispersion relations as,
\be
\omega_\pm = c_{\pm} k_\pm \, ,
\ee
{}from where we obtain the group velocities as,
\be
c_\pm = \frac{-\left(a_+ - a_-\right) \pm \sqrt{(a_+ + a_-)^2 + 4 d^2}}{2\beta^2}
\ee
It should be observed that
\be
c_+ \, c_- = - \frac 1{\beta^2}\, ,
\ee
showing that these are propagating modes with opposite chiralities.
It should also be noticed that in the limit of vanishing deformation, $d\to 0$, 
\be
\label{cvelocity2}
c_\pm \to \pm \frac 1{a_\pm(\theta)}\, ,
\ee
as expected.

To implement the soldering we follow the earlier procedure, where the Noether current is
\bn
\label{current2}
J =  &2&\left[\left(a_+ \lambda_+ + d\, \lambda_-\right)\dot\varphi_+ - \left(a_-\lambda_- - d\lambda_+\right)\dot\varphi_-\right.\nonumber\\
&-& \left. \left(\lambda_+\varphi_+'+\lambda_-\varphi_- '\right)\right]\, .
\en
As before, the second-iterated action
becomes gauge-invariant if the following condition over the soldering parameters holds,
\be
a_+\, \lambda_+^2 - a_-\, \lambda_-^2 + 2\, d\, \lambda_+ \lambda_- =0\, ,
\ee
that combines both earlier conditions, Eqs.(\ref{6}) and (\ref{condition2}).
This constraint may be written as a pair of conditions as,
\be
\l{consistency}
a_+ \frac {\lambda_+}{\lambda_-} + d = a_- \frac {\lambda_-}{\lambda_+} - d = \beta
\ee
which, in matrix form, read
\be
\pmatrix{a_+(\theta)  & {d(\theta) - \beta} \cr
	{d(\theta) + \beta}  & {- a_-(\theta)} \cr}\pmatrix{\lambda_+ \cr -\lambda_- \cr} = 0\, .
\ee
The condition for the non triviality solution for $\vec \lambda$ vector demands $\beta$ to satisfy (\ref{beta}), consistent with the dispersion relation solution.
Using (\ref{consistency}) we may write the Noether current as
\be
\l{noether2}
J = 2\left[\beta\left(\lambda_- \dot\varphi_+ - \lambda_+ \dot\varphi_-\right)
- \left(\lambda_- \varphi_-' + \lambda_+ \varphi_+'\right) \right]\, .
\ee

Elimination of the soldering field $B$ leads to an effective action as
\be
{\cal L}_{eff}\,=\,{\cal L}_0\,+\, \frac 1{4\left(\lambda_+^2 + \lambda_-^2\right)}\,{ J}^2\, .
\ee
This effective action may be cast in a more interesting form with the introduction of the soldered doublet
\be
\Phi = \frac{\sqrt{2}\beta}{\sqrt{\lambda_+^2 + \lambda_-^2}}\left(\lambda_- \varphi_+ - \lambda_+ \varphi_-\right)\, .
\ee
Indeed, using (\ref{noether2}) together with solution for the soldering parameters above, the soldered action becomes
\be
{\cal L}_{eff} = \frac 12 {\dot\Phi}^2 + \frac 1{2\beta^2} {\Phi'}^2 + \frac{a_+ - a_-}{2\beta^2}\dot\Phi \Phi'.
\ee
Consistency with previous results are obtained by adopting the the previous normalization (\ref{normalization}) and scaling the doublet as $\Phi\to\beta\Phi$. It should be noticed that the dispersion relation of this theory match that of the chiral bosons, as expected, leading to modes with the same group velocity as before. It should also be observe that the mix Kostelecky term depends explicitly on the presence of the birefringence since in general $a_+ - a_-\neq 0$. This has extended the situation considered in the previous section.

\section{Final Remarks}

Taking in account the current technology and the astonishing energy scales involved, setting up an experimental search for new physics at the Planck scale is pointless. For the symmetries, however, the situation is quite distinct since some high-energy theories may lead to violations which hold exactly in the low-energy effective model. Therefore there exists hope that Lorentz symmetry violations can be tested due to the extremely high precision of today's technology.

With this motivation in mind we have studied issues of Lorentz symmetry violation in a $D=2$ model of noncommutative chiral bosons using the soldering formalism.  Within the noncommutative field theory put forward by \cite{CCGM}, we have considered the effects of fusing chiral components under different conditions that includes deformations of the symplectic structure and vacuum birefringence. Our results are consistent with those obtained in \cite{DGMJ}, for the former instance and with \cite{CK} for the latter. To effect this study it became necessary to extend the concept of soldering - besides the mutual cancellation of chiral obstructions to gauge symmetry, a new condition in the form of a constraint has to be imposed over the parameters of soldering. This condition was shown to lead to a real one-parameter solution that generalizes the soldering. 

The scenario without birefringence but with a deformed symplectic structure, proposed in \cite{DGMJ} was considered first and shown to lead to a physical situation where a preferred-frame effect was avoided at the cost, in general, of changing the {\it speed of light} for the resulting scalar composite.  However, a new class of solutions not disclosed in \cite{DGMJ} was found that, despite the deformation, keeps unchanged the velocity of propagation for the scalar mode.  A physical interpretation was offered showing that this class corresponds to the imposition of a generalized chiral constraint over a pair of scalar fields, extending the analogous situation where the one-field chiral constraint $\pi = \pm \,\varphi '$ imposed over a single scalar produces the action for the chiral boson.

Finally the physically more motivated situation was considered in Section III where issues of preferred-frame effects associated to the birefringence of the $D=2$ vacuum were considered. We found that the soldering of the chiral modes leads directly to the scalar sector of the extended Standard Model proposed by Kostelecky. This shows the consistency of the soldering approach with earlier studies in the limiting scalar sector of the Standard Model. It would be interesting to consider next the interference like effects of soldering in models with non-commutative fields in dimensions higher than $D=2$.


We would like to thank CAPES/PROCAD, CNPq/PRONEX/FAPESQ and FAPERJ for financial support. One of us (CW) would like to thank the Department of Physics at USACH-Chile for their hospitality in the earlier stages of this researches. He also thanks M. Berger and J. Gamboa for interesting discussions.


\begin{thebibliography}{99}

\bibitem{CCGM}  J. Carmona, J. L. Cort\'es, J. Gamboa and F. M\'endes, JHEP 03 (2003) 058; {\it ibid} Phys. Lett. B 565 (2003) 222.

\bibitem{DGMJ}  A. Das, J. Gamboa, F. M\'endes and J. L\'opes-Sarri\'on, JHEP 0405:022,2004.

\bibitem{string} A. Connes, M. Douglas and A. Schwarz, JHEP 02 (1998) 003. For recent reviews: N. A. Nekrasov, hep-th/0011095; A. Konechny and A. Schwarz, hep-th/0012145; J. A. Harvey, hep-th/0102076; N. Seiberg and E. Witten, JHEP 09 (1999) 032; G. Mandanici and A. Marcian\`o, JHEP 0409:040,2004.

\bibitem{astrobservation} G. Amelino-Camelia and T. Piran, Phys.Rev.D64 (2001) 036005; T. Jacobson,
S. Liberati, D. Mattingly, Phys. Rev. D 66 (2002) 081302(R); T. J. Konopka and S. A. Major, New J. Phys. 4 (2002) 57; R. Aloisio, P. Blasi, A. Galante, A.F. Grillo, Astropart. Phys. 20 (2003) 369; O.Bertolami, C.S. Carvalho, Phys. Rev. D61 (2000) 103002.

\bibitem{CK} D. Colladay and V. A. Kostelecky, Phys. Rev. D55 (1997) 6760; D58 (1998) 116002; V. A. Kostelecky and R. Potting, Phys. Rev. D51 (1995) 3923.

\bibitem{quantumgravity} G. Amelino-Camelia, J. Ellis, N. E. Mavromatos, D. V. Nanopoulos and S. Sarkar, Nature 393 (1998) 763; R. Gambini and J. Pullin, Phys. Rev. D59 (1999) 124021; J. Alfaro, H. A. Morales-Tecotl and L. F. Urrutia, Phys. Rev. Lett. 84 (2000) 2318; Phys. Rev. D65 (2002) 103509; J. Alfaro and G. Palma, Phys. Rev. D65 (2002) 103516.

\bibitem{highenergy} D. Colladay and V. A. Kostelecky, Phys.Lett. B511 (2001) 209; V. A. Kostelecky and R. Lehnert, Phys. Rev. D63 (2001) 065008; R. Bluhm and V. A. Kostelecky, Phys. Rev. Lett. 84 (2000) 1381; R. Jackiw and V. A. Kostelecky, Phys. Rev. Lett. 82 (1999) 3572; O. Bertolami, Class.Quant.Grav.14:2785-2791,1997.


\bibitem{Chadha:1982qq}
  S.~Chadha and H.~B.~Nielsen,
  Nucl.\ Phys.\ B { 217} (1983) 125.

\bibitem{np}   H. B. Nilsen and I. Picek, Phys. Lett. B 114 (1982) 141; Nucl Phys. B 211 (1983) 269.

\bibitem{klinkhamer}   F. R. Klinkhamer, Phys. Rev. D 66 (2002) 047701.

\bibitem{klp}   V. A. Kostelecky, R. Lehnert and M. J. Perry, Phys. Rev. D 68 (2003) 123511.

\bibitem{qednc} S. M. Carroll, J. A. Harvey, V. A. Kostelecky, C. D. Lane and T. Okamoto, Phys. Rev. Lett. 87 (2001) 141601; O. Bertolami and L. Guisado, JHEP 0312:013,2003.

\bibitem{VSL} S. Alexander and J. Magueijo, Noncommutative geometry as a realization of varying speed of light cosmology; hep-th/0104093.

\bibitem{solda} M. Stone, Phys. Rev. Lett. 63 (1989) 731; Nucl. Phys. B 327
(1989) 399; Report no. ILL-23/89 (unpublished); D. Depireux, S. J. Gates Jr.
and Q-Han Park, Phys. Lett. B 224 (1989); E. Witten, Commun. Math. Phys. 144
(1992) 189; E. M. C. Abreu, R. Banerjee and C. Wotzasek, Nucl. Phys. B 509
(1998) 519; R. Amorim, A. Das and C. Wotzasek, Phys. Rev. D 53 (1996) 5810;
S.~Ghosh,
  Phys.\ Lett.\ B {579} (2004) 377.

\bibitem{capitulo}  C. Wotzasek, ``Soldering Formalism: theory and applications", hep-th/9806005;
E. M. C. Abreu and C. Wotzasek, ``Topics on the Quantum Dynamics of Chiral Bosons", to appear in ``Progress in Boson Research" Nova Eds., NY, USA, hep-th/0410019.
  
\bibitem{siegel} W. Siegel, Nucl. Phys. B 238 (1984) 307.

\bibitem{florjack} R. Floreanini and R. Jackiw, Phys. Rev. Lett. 59 (1987) 1873.

\bibitem{abreuwotz}  E. M. C. Abreu and C. Wotzasek, Phys. Rev. D 58, 101701(R) (1998).

\bibitem{hull}   C. M. Hull, Phys. Lett. B 206 (1988) 234; B212 (1988) 437.

\bibitem{scalar} O. Bertolami, R. Lehnert, R. Potting, and A. Ribeiro, Phys.Rev.D69:083513,2004.

\bibitem{lehnert}  R. Lehnert, Phys. Rev. D 68 (2003) 085003.

\bibitem{km}  V. A. Kostelecky and M. Mewes, Phys. Rev. D 66 (2002) 056005. 



\end{thebibliography}
\end{document}